\documentclass[aps,prl,twocolumn,showpacs,superscriptaddress]{revtex4}

\usepackage{graphicx}

\newcommand{\is}[2]{$^{#2}$#1}

\begin{document}

\title{Direct mass measurements beyond the proton drip-line}

\author{C.~Rauth}
\email[Electronic address: ]{c.rauth@gsi.de}
\author{D.~Ackermann}
\affiliation{GSI, Planckstr.~1, 64291 Darmstadt, Germany}
\author{K.~Blaum}
\affiliation{Institut f\"ur Physik, Johannes Gutenberg-Universit\"at, 55099 Mainz, Germany}
\author{M.~Block}
\altaffiliation[Present address: ]{NSCL, MSU, 1 Cyclotron, East Lansing, 48824 MI, USA.}
\affiliation{GSI, Planckstr.~1, 64291 Darmstadt, Germany}
\author{A.~Chaudhuri}
\affiliation{Institut f\"ur Physik, Ernst-Moritz-Arndt-Universit\"at, 17489 Greifswald, Germany}
\author{S.~Eliseev}
\affiliation{GSI, Planckstr.~1, 64291 Darmstadt, Germany}
\affiliation{St.~Petersburg Nuclear Physics Institute, Gatchina 188300, Russia}
\author{R.~Ferrer}
\affiliation{Institut f\"ur Physik, Johannes Gutenberg-Universit\"at, 55099 Mainz, Germany}
\author{D.~Habs}
\affiliation{Department f\"ur Physik, Ludwig-Maximilians-Universit\"at M\"unchen, 85748 Garching, Germany}
\author{F.~Herfurth}
\affiliation{GSI, Planckstr.~1, 64291 Darmstadt, Germany}
\author{F.~P.~He\ss{}berger}
\affiliation{GSI, Planckstr.~1, 64291 Darmstadt, Germany}
\author{S.~Hofmann}
\affiliation{GSI, Planckstr.~1, 64291 Darmstadt, Germany}
\affiliation{Institut f\"ur Physik, Johann Wolfgang Goethe-Universit\"at Frankfurt, 60438 Frankfurt, Germany}
\author{H.-J.~Kluge}
\affiliation{GSI, Planckstr.~1, 64291 Darmstadt, Germany}
\author{G.~Maero}
\affiliation{GSI, Planckstr.~1, 64291 Darmstadt, Germany}
\author{A.~Mart\'in}
\affiliation{GSI, Planckstr.~1, 64291 Darmstadt, Germany}
\author{G.~Marx}
\affiliation{Institut f\"ur Physik, Ernst-Moritz-Arndt-Universit\"at, 17489 Greifswald, Germany}
\author{M.~Mukherjee}
\altaffiliation[Present address: ]{Universit\"at Innsbruck, 6020 Innsbruck, Austria}
\affiliation{GSI, Planckstr.~1, 64291 Darmstadt, Germany}
\author{J.~B.~Neumayr}
\affiliation{Department f\"ur Physik, Ludwig-Maximilians-Universit\"at M\"unchen, 85748 Garching, Germany}
\author{W.~R.~Pla\ss{}}
\affiliation{II.~Physikalisches Institut, Justus-Liebig-Universit\"at, 35392 Gie\ss{}en, Germany}
\author{W.~Quint}
\affiliation{GSI, Planckstr.~1, 64291 Darmstadt, Germany}
\author{S.~Rahaman}
\altaffiliation[Present address: ]{University of Jyv\"askyl\"a, P.O.~Box 35 (YFL), 40014 Jyv\"askyl\"a, Finland}
\affiliation{GSI, Planckstr.~1, 64291 Darmstadt, Germany}
\author{D.~Rodr\'iguez}
\altaffiliation[Present address: ]{Departamento de f\'isica aplicada, University of Huelva, 21071 Huelva, Spain}
\affiliation{IN2P3, LPC-ENSICAEN, 6 Bd.~Marechal Juin, 14050 Caen Cedex, France}
\author{C.~Scheidenberger}
\affiliation{GSI, Planckstr.~1, 64291 Darmstadt, Germany}
\author{L.~Schweikhard}
\affiliation{Institut f\"ur Physik, Ernst-Moritz-Arndt-Universit\"at, 17489 Greifswald, Germany}
\author{P.~G.~Thirolf}
\affiliation{Department f\"ur Physik, Ludwig-Maximilians-Universit\"at M\"unchen, 85748 Garching, Germany}
\author{G.~Vorobjev}
\affiliation{GSI, Planckstr.~1, 64291 Darmstadt, Germany}
\affiliation{St.~Petersburg Nuclear Physics Institute, Gatchina 188300, Russia}
\author{C.~Weber}
\altaffiliation[Present address: ]{University of Jyv\"askyl\"a, P.O.~Box 35 (YFL), 40014 Jyv\"askyl\"a, Finland}
\affiliation{GSI, Planckstr.~1, 64291 Darmstadt, Germany}
\affiliation{Institut f\"ur Physik, Johannes Gutenberg-Universit\"at, 55099 Mainz, Germany}
\author{Z.~Di}
\affiliation{II.~Physikalisches Institut, Justus-Liebig-Universit\"at, 35392 Gie\ss{}en, Germany}

\date{\today}

\begin{abstract}
First on-line mass measurements were performed at the SHIPTRAP Penning trap mass spectrometer. The masses of 18 neutron-deficient isotopes in the terbium-to-thulium region produced in fusion-evaporation reactions were determined with relative uncertainties of about $7\cdot 10^{-8}$, nine of them for the first time. Four nuclides ($^{144, 145}$Ho and $^{147, 148}$Tm) were found to be proton-unbound. The implication of the results on the location of the proton drip-line is discussed by analyzing the one-proton separation energies.
\end{abstract}

\pacs{23.50.+z, 07.75.+h, 21.10.Dr, 27.60.+j}

\maketitle
The origin and the abundance pattern of the chemical elements in our solar system is one of the major unsolved problems in modern physics. The paths of the stellar production mechanisms for heavy elements depend strongly on input data from nuclear physics like $Q$-values, branching ratios and decay constants. Here, precise mass measurements of exotic nuclei can serve as an essential source of information, especially where spectroscopic investigations are not possible \cite{Schatz_RpProcess,Kluge_SpecialEdition,Lunney_TrendsInMasses,Blaum_Review}.\par
An important topic in this context is the location of the borderlines of nuclear stability. While the neutron drip-line is experimentally still only reached for very light elements, the proton drip-line was accessed up to heavy elements like protactinium \cite{Novikov_NPA}. The first ground state proton emitter, \is{Lu}{151}, was discovered at GSI in 1981 \cite{Sigurd_Lu151}. Other proton-decaying nuclides like \is{Tm}{147} were observed shortly after \cite{Klepper_Tm147}. Whereas the detection of proton decay is of course sufficient proof of the unbound character of the emitting state, the inverse relation does not necessarily hold: The proton radioactivity of a proton-unbound nuclide may well be too weak to be detected, especially in close vicinity to the proton drip-line. This is due to two effects. First, the $Q$-value of direct proton decay $Q_p$ strongly affects the decay rate. A small $Q_p$ value results in a very low decay rate, i.e., a very long partial half-life, and thus a negligibly small branching ratio compared to the competing $\beta$ decay. Second, it is experimentally very challenging to discriminate the low-energy proton against the background of $\beta$ decay positrons. For these reasons proton emitters are generally found only at some distance from the proton drip-line and can not be used to delineate its location precisely. This letter shows for the first time that precise mass measurements can overcome these problems through a determination of the $Q_p$ value by measuring the masses of the mother and the daughter nuclei of the direct proton decay.\par
The SHIPTRAP facility \cite{Michael_EPJAPaper} was set up at GSI Darmstadt to perform precision experiments with radioactive nuclides produced in fusion-evaporation reactions at the velocity filter SHIP \cite{SHIPReview}. These experiments aim for investigations of the nuclear or atomic structure by means of trap-assisted decay spectroscopy, mass or laser spectroscopy, and chemical reaction studies. In the first stage SHIPTRAP focuses on accurate mass measurements with a Penning trap mass spectrometer. The location behind SHIP offers the opportunity to investigate radioactive ions from medium-heavy up to transuranium elements. This letter addresses the first results obtained using SHIPTRAP, which are mass measurements of neutron-deficient nuclides in the rare-earth region ($A\approx 146$). These measurements are particularly interesting because of the proximity to the proton drip-line and the occurrence of ground-state proton radioactivity.\par
\begin{figure*}[t]
  \includegraphics[width=.8\linewidth]{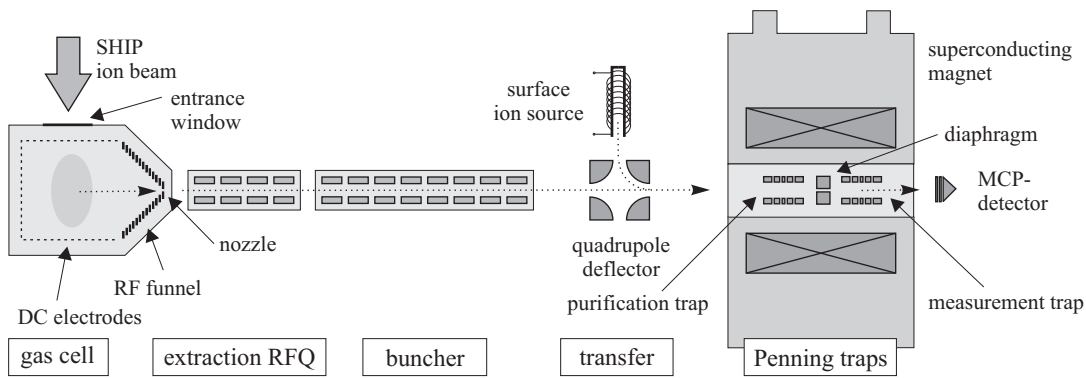}
  \caption[Schematic overview of the SHIPTRAP experiment]{\label{fig:SetUp}Schematic overview of the SHIPTRAP experiment. The radioactive ions delivered by SHIP are stopped (gas cell section), accumulated and cooled (buncher section) and transferred (transfer section) into the 7\,T superconducting magnet (trap section). After isobaric separation (purification trap) the cyclotron frequency is determined (measurement trap).}
\end{figure*}
A stable-isotope beam accelerated to energies of about 4--5\,MeV/u impinges on a thin target foil to produce radioactive ions via fusion-evaporation reactions. The slower reaction products with energies of typically few 100\,keV/u are kinematically separated from the faster primary beam and transferred to SHIPTRAP. A schematic overview of the setup is shown in Fig.~\ref{fig:SetUp}. The ions enter a helium-filled buffer-gas cell \cite{Juergen_GasCell} through a few $\mu$m thin metal foil. While the main energy loss happens in the window foil the ions are finally thermalized inside the cell by collisions with the helium buffer gas. A combination of electrical DC and radio frequency (RF) fields drags the ions towards the nozzle region, where they are swept out by a supersonic gas jet into the extraction RF quadrupole (RFQ) that acts as a differential pumping line and ion guide. The subsequent section consists of a helium-filled RFQ cooler and buncher, which accumulates the ions and enhances the injection efficiency into the Penning traps by decreasing the emittance of the ion beam \cite{Frank_Buncher}. The following section contains ion-optical elements to deflect and focus the ion beam into the magnet. A quadrupole deflector is used to quickly switch between the radioactive ions from the gas cell and stable ions from a dedicated reference ion source. For the latter purpose a surface ion source containing a mixture of alkali elements (potassium, rubidium and cesium) or alternatively a carbon-cluster ion source \cite{Ankur_Paper} is used. In the last section a 7\,T superconducting magnet houses two cylindrical Penning traps in two homogeneous-field regions. In the first Penning trap (purification trap) a mass-selective buffer-gas technique \cite{Savard_BufferGasCooling} is applied to select one nuclide, which is transferred through a 3\,mm diaphragm into the second Penning trap (measurement trap). Here, the mass determination is performed by measuring the cyclotron frequency
\begin{figure}[b]
  \includegraphics[width=.9\linewidth]{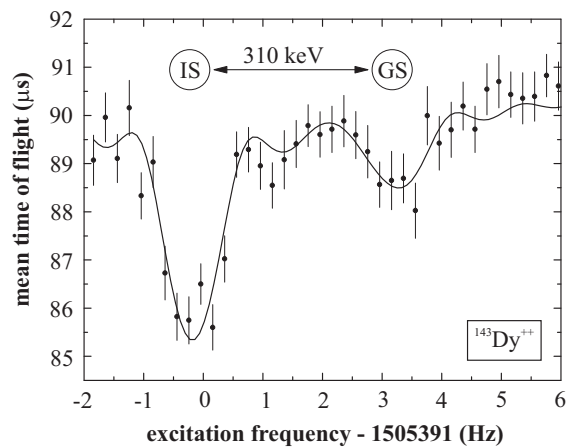}
  \caption[Resonance of \is{Dy}{143}$^{++}$]{\label{fig:Dy143Res}Cyclotron resonance of \is{Dy}{143}$^{++}$. An excitation time of 900\,ms leading to a mass resolving power of $1.2\cdot 10^6$ was used to separate the ground state (right) from the isomeric state (left).}
\end{figure}
\begin{equation}
	\nu_c=\frac{1}{2\pi}\frac{ze}{m_{ion}}B\quad,
	\label{equ:CyclFrequ}
\end{equation}
%
which is the frequency with which an ion of mass $m_{ion}$ and charge state $z$ rotates in a magnetic field of strength $B$. This frequency is measured by means of the time-of-flight ion-cyclotron-resonance (TOF-ICR) method \cite{Graeff_TOFICR}. The magnetic field strength is calibrated by measuring the cyclotron frequency $\nu_{ref}$ of a reference ion with a well-known mass $m_{ref}$. Taking the frequency ratio $r=\nu_{ref}/\nu_c$ the mass of the atomic nuclide can finally be calculated as 
\begin{equation}
	m=\frac{z}{z_{ref}}r\,(m_{ref}-z_{ref}m_e)+z\,m_e\quad,
	\label{equ:MassCalc}
\end{equation}
where $m_e$ is the electron mass and $z_{ref}$ the charge state of the reference ion. The binding energies of some eV for the outer electrons can be neglected for a mass uncertainty above 1\,keV. A detailed description of the complete setup and the measurement procedure can be found in \cite{Michael_NIMPaper}.\par
The data presented here were obtained in two experiments performed in October 2005 and December 2005. A \is{Ni}{58}$^{14+}$ beam with an energy of 4.36\,MeV/u and 4.60\,MeV/u hit a 0.626\,mg/cm$^2$ thick \is{Mo}{92} foil to produce neutron-deficient isotopes in the terbium-to-thulium region. The primary beam intensity was around 200\,particle-nA.\par
For the first identification of the radionuclides in the purification trap an excitation time of 200\,ms yielded a mass resolving power of about 50,000. This was sufficient for a clean selection of the ion of interest before the transfer to the measurement trap. The excitation time here was usually 900\,ms leading to a resolving power of $1.2\cdot 10^6$. In total, the masses of 18 short-lived nuclides were measured as summarized in Table~\ref{tab:MassExcess} compared to the previous literature values \cite{AME2003}. The mass values of nine nuclides had been measured before, mainly at the Experimental Storage Ring (ESR) of GSI \cite{Litvinov_ESRMeas} and by the on-line Penning trap mass spectrometer ISOLTRAP \cite{Didi_ISOLTRAP}, while the other nine had been only estimated by extrapolation of the mass surface. In comparison with the previous data the new results mainly agree with the measured ones but slightly deviate from the estimates. The only significant deviation was found for \is{Ho}{147}, its origin presently not being understood. The analysis of the data yielded a relative mass uncertainty of $6.8\cdot 10^{-8}$ in average \cite{MyEPJAPaper}.\par
\begin{table}[tb]
  \caption[Results of the atomic mass evaluation]{\label{tab:MassExcess}Results of the mass measurements. The half-lives T$_{1/2}$ are listed according to \cite{AME2003}. Column three and four give the mass excesses of the data obtained in this work (M$_{\mathrm{exp}}$) and the previous literature values (M$_{\mathrm{AME}}$) \cite{AME2003}. Estimated masses are marked with $^\#$. Column four shows the differences between the experimental and the previous data (M$_{\mathrm{exp}} - $M$_{\mathrm{AME}}$).}
  \begin{ruledtabular}
  \begin{tabular}{llllr}
Nuclide & T$_{1/2}$ & M$_{\mathrm{exp}}$(keV) & M$_{\mathrm{AME}}$(keV) & Diff.(keV) \\ \hline
\is{Tb}{143}	&	12\,s	&	-60420(50)	&	-60430(60)	&	10(78)	\\
\is{Tb}{147}	&	1.64\,h	&	-70740(11)	&	-70752(12)	&	12(16)	\\
\is{Dy}{143}	&	5.6\,s	&	-52169(13)	&	-52320(200)$^\#$	&	150(200)	\\
\is{Dy}{144}	&	9.1\,s	&	-56570.1(7.1)	&	-56580(30)	&	10(31)	\\
\is{Dy}{145}	&	9.5\,s	&	-58242.6(6.5)	&	-58290(50)	&	47(50)	\\
\is{Dy}{146}	&	33.2\,s	&	-62554.7(6.7)	&	-62554(27)	&	1(28)	\\
\is{Dy}{147}	&	40\,s	&	-64197.9(8.8)	&	-64188(20)	&	-10(22)	\\
\is{Dy}{148}	&	3.3\,min	&	-67861(13)	&	-67859(11)	&	-2(17)	\\
\is{Ho}{144}	&	700\,ms	&	-44609.5(8.5)	&	-45200(300)$^\#$	&	590(300)	\\
\is{Ho}{145}	&	2.4\,s	&	-49120.1(7.5)	&	-49180(300)$^\#$	&	60(300)	\\
\is{Ho}{146}	&	3.6\,s	&	-51238.2(6.6)	&	-51570(200)$^\#$	&	330(200)	\\
\is{Ho}{147}	&	5.8\,s	&	-55757.1(5.0)	&	-55837(28)	&	80(28)	\\
\is{Ho}{148}	&	2.2\,s	&	-57990(80)	&	-58020(130)	&	30(150)	\\
\is{Er}{146}	&	1.7\,s	&	-44325.0(8.6)	&	-44710(300)$^\#$	&	390(300)	\\
\is{Er}{147}	&	2.5\,s	&	-46610(40)	&	-47050(300)$^\#$	&	440(300)	\\
\is{Er}{148}	&	4.6\,s	&	-51479(10)	&	-51650(200)$^\#$	&	170(200)	\\
\is{Tm}{147}	&	580\,ms	&	-35969.8(9.9)	&	-36370(300)$^\#$	&	400(300)	\\
\is{Tm}{148}	&	700\,ms	&	-38765(10)	&	-39270(400)$^\#$	&	500(400)	\\
  \end{tabular}
  \end{ruledtabular}
\end{table}
The high resolving power allowed us to separate the first isomeric states of \is{Dy}{143} (see Fig.~\ref{fig:Dy143Res}) and \is{Dy}{147} with excitation energies of about 310\,keV and 750\,keV \cite{AME2003}, respectively. The individual frequency ratios for the ground and the isomeric states were both used to determine the mass of the corresponding ground state. The measurement of \is{Tm}{147}, with a half-life of 580\,ms and a production cross-section of only 100--200\,$\mu$b \cite{Klepper_Tm147,Seweryniak_Tm147}, allowed us to estimate the overall efficiency of SHIPTRAP to $2\cdot10^{-4}$.\par 

The proton separation energy $S_p$, defined as
\begin{eqnarray}
  S_p & = &  B(Z,N)-B(Z-1,N) \nonumber \\
      & = & -M(Z,N)+M(Z-1,N)+M_H \quad ,
\end{eqnarray}
where $M_H$ is the atomic mass of hydrogen, allows one to distinguish between the proton-bound ($S_p>0$) and the proton-unbound ($S_p<0$) nuclei or in other words to determine the proton drip-line. From the present data the $S_p$ values of four nuclides (\is{Ho}{144, 145} and \is{Tm}{147, 148}) were determined to be negative. Table~\ref{tab:Sp} shows the results in comparison with the previous values. Only \is{Tm}{147} is a known and measured proton emitter \cite{Klepper_Tm147,Sellin_Tm147} while for the other three nuclides the $S_p$ values had prior to this work only been deduced from mass estimates. However, due to the large uncertainties of these estimates, they did not allow for an unambiguous assignment of the sign of $S_p$ and therefore of the position of the drip-line.\par
\begin{figure}[t]
\includegraphics[width=.9\linewidth]{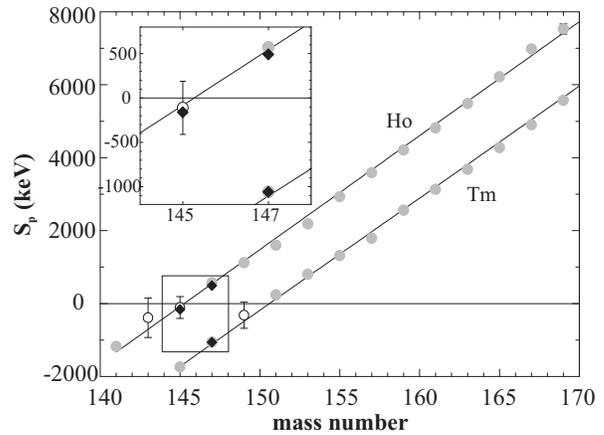}
\caption[Proton separation energies]{\label{fig:Sp}Proton separation energies versus mass number for the odd-$Z$ elements holmium ($Z=67$) and thulium ($Z=69$). Only the even-$N$ isotopes are plotted. The circles show the previous measurements (full gray circles) and estimations (empty circles) taken from \cite{AME2003}. The data from this work are represented by black diamonds. Also shown are linear fits to the data. The inner box enlarges the interesting region around the drip-line.}
\end{figure}

The determination of the proton drip-line on the basis of experimental data is visualized in Fig.~\ref{fig:Sp}. Here the $S_p$ values for the two elements holmium and thulium are plotted for their even-$N$ isotopes only, thus avoiding the odd-even staggering due to the pairing energy. Three data points from the present work were included. The new $S_p$ value obtained for \is{Tm}{147} agrees well with that determined previously \cite{Sellin_Tm147}. The measurement of \is{Ho}{147} slightly shifts the previous result to a lower separation energy. The value of \is{Ho}{145} was never measured before and the measurements clearly show that \is{Ho}{145} is proton-unbound. All data agree with the expected linear trend and confirm the location of the drip-lines: between $A=145$ and $A=147$ for holmium and $A=149$ and $A=151$ for thulium.\par
The question whether the three new proton-unbound nuclides are accessible to investigations of direct proton decay can be answered by looking at the partial half-lives of this decay mode. In a simple, semi-empirical approach applying the WKB approximation the proton decay can be described similarly to the $\alpha$ decay \cite{Sigurd_ProtonRadioactivity}. The decay constant
\begin{equation}
  \lambda_p = \frac{\ln 2}{T_p} = \nu_0 \cdot e^{-2C_p}
  \label{eqn:lambda}
\end{equation}
is in this picture the product of a frequency factor $\nu_0$ and an exponential transmission term. The factor $\nu_0$ is in first order the inverse of the characteristic nuclear time and mainly defined by the velocity of the proton and the radius of the nucleus. In the investigated region of the nuclear chart $\nu_0$ has a typical value of $6\cdot 10^{22}$\,Hz \cite{Sigurd_ProtonRadioactivity} and is more or less constant. The transmission term, usually called Gamow factor, describes the tunneling probability through the Coulomb and centrifugal potential barriers. The exponent $C_p$ is the integral over the `forbidden' region of the potential. It depends mainly on the energy and angular momentum of the proton and can be calculated numerically. An angular momentum of 5 was estimated by comparing the calculated half-life of \is{Tm}{147} with the measured one. All calculated half-lives are shown in the last column of Table~\ref{tab:Sp}. Only for \is{Tm}{147} the proton decay is an observable decay channel. While for \is{Tm}{148} it is experimentally very difficult to discriminate the proton from the $\beta$ background, \is{Ho}{144} and \is{Ho}{145} are surely far out of reach for the study of proton decay as their partial half-lives for this decay mode are longer than the age of the universe.\par
\begin{table}[tb]
  \caption[Nuclides with negative $S_p$ value]{\label{tab:Sp}List of nuclides with negative $S_p$ value. The partial half-lives for proton decay $T_p$ are given in the last column according to Eq.~(\ref{eqn:lambda}).}
  \begin{ruledtabular}
  \begin{tabular}{lrlc}
Nuclide&\multicolumn{2}{c}{$S_p$ (keV)}&$T_p$ (s) \\
&This work&Literature&\\\hline
\is{Ho}{144}	&	-271(16) & -160(360) & $10^{25}$ \\
\is{Ho}{145}	&	-161(10) & -110(300) & $10^{40}$ \\
\is{Tm}{147}	&	-1066(13) & -1058(3)    & 3         \\
\is{Tm}{148}	&	-560(40) & -490(500) & $10^{11}$ \\
  \end{tabular}
  \end{ruledtabular}
\end{table}
In summary, we present the first on-line mass measurements at the Penning trap spectrometer SHIPTRAP. The masses of 18 radionuclides in the neutron-deficient terbium-to-thulium region were measured, nine of them for the first time. The results show generally good agreement with the previously measured or estimated data. By analyzing the proton separation energies the new mass values were used to pin down the location of the proton drip-line in this region unambiguously. Simple model calculations of the partial half-lives show that direct measurements of the proton decay are not feasible or in the case of \is{Tm}{148} at least extremely ambitious.
\begin{acknowledgments}
The authors would like to thank G.~Audi, Y.~Novikov and E.~Roeckl for their valuable support and fruitful discussions. This project was supported by the BMBF and the EC under contract number RII3-CT-2004-506065 (EURONS/JRA11/TRAPSPEC).
\end{acknowledgments}
%
%

\begin{thebibliography}{19}
\expandafter\ifx\csname natexlab\endcsname\relax\def\natexlab#1{#1}\fi
\expandafter\ifx\csname bibnamefont\endcsname\relax
  \def\bibnamefont#1{#1}\fi
\expandafter\ifx\csname bibfnamefont\endcsname\relax
  \def\bibfnamefont#1{#1}\fi
\expandafter\ifx\csname citenamefont\endcsname\relax
  \def\citenamefont#1{#1}\fi
\expandafter\ifx\csname url\endcsname\relax
  \def\url#1{\texttt{#1}}\fi
\expandafter\ifx\csname urlprefix\endcsname\relax\def\urlprefix{URL }\fi
\providecommand{\bibinfo}[2]{#2}
\providecommand{\eprint}[2][]{\url{#2}}

\bibitem[{\citenamefont{Schatz et~al.}(2001)\citenamefont{Schatz et~al.}}]{Schatz_RpProcess}
\bibinfo{author}{\bibfnamefont{H.}~\bibnamefont{Schatz}}
  \bibnamefont{et~al.},
  \bibinfo{journal}{Phys. Rev. Lett.} \textbf{\bibinfo{volume}{86}},
  \bibinfo{pages}{3471} (\bibinfo{year}{2001}).
  
\bibitem[{\citenamefont{Schweikhard and Bollen}(2006)}]{Kluge_SpecialEdition}
\bibinfo{author}{\bibfnamefont{L.}~\bibnamefont{Schweikhard}} \bibnamefont{and}
  \bibinfo{author}{\bibfnamefont{G.}~\bibnamefont{Bollen}}, eds., 
  \bibinfo{journal}{Int. J. Mass Spectrom.} \textbf{\bibinfo{volume}{251}},
  \bibinfo{pages}{85} (\bibinfo{year}{2006}).

\bibitem[{\citenamefont{Lunney et~al.}(2003)\citenamefont{Lunney, Pearson, and
  Thibault}}]{Lunney_TrendsInMasses}
\bibinfo{author}{\bibfnamefont{D.}~\bibnamefont{Lunney}},
  \bibinfo{author}{\bibfnamefont{J.~M.}~\bibnamefont{Pearson}},
  \bibnamefont{and} \bibinfo{author}{\bibfnamefont{C.}~\bibnamefont{Thibault}},
  \bibinfo{journal}{Rev. Mod. Phys.} \textbf{\bibinfo{volume}{75}},
  \bibinfo{pages}{1021} (\bibinfo{year}{2003}).

\bibitem[{\citenamefont{Blaum}(2006)}]{Blaum_Review}
\bibinfo{author}{\bibfnamefont{K.}~\bibnamefont{Blaum}},
  \bibinfo{journal}{Phys. Rep.} \textbf{\bibinfo{volume}{425}},
  \bibinfo{pages}{1} (\bibinfo{year}{2006}).

\bibitem[{\citenamefont{Novikov et~al.}(2002)\citenamefont{Novikov et~al.}}]{Novikov_NPA}
\bibinfo{author}{\bibfnamefont{Y.~N.}~\bibnamefont{Novikov}}
  \bibnamefont{et~al.}, \bibinfo{journal}{Nucl. Phys. A}
  \textbf{\bibinfo{volume}{697}}, \bibinfo{pages}{92} (\bibinfo{year}{2002}).

\bibitem[{\citenamefont{Hofmann et~al.}(1982)\citenamefont{Hofmann et~al.}}]{Sigurd_Lu151}
\bibinfo{author}{\bibfnamefont{S.}~\bibnamefont{Hofmann}}
  \bibnamefont{et~al.},
  \bibinfo{journal}{Z. Phys. A} \textbf{\bibinfo{volume}{305}},
  \bibinfo{pages}{111} (\bibinfo{year}{1982}).

\bibitem[{\citenamefont{Klepper et~al.}(1982)\citenamefont{Klepper et~al.}}]{Klepper_Tm147}
\bibinfo{author}{\bibfnamefont{O.}~\bibnamefont{Klepper}}
  \bibnamefont{et~al.},
  \bibinfo{journal}{Z. Phys. A} \textbf{\bibinfo{volume}{305}},
  \bibinfo{pages}{125} (\bibinfo{year}{1982}).

\bibitem[{\citenamefont{Block et~al.}(2005)\citenamefont{Block
  et~al.}}]{Michael_EPJAPaper}
\bibinfo{author}{\bibfnamefont{M.}~\bibnamefont{Block}}
  \bibnamefont{et~al.}, \bibinfo{journal}{Eur. Phys. J. A}
  \textbf{\bibinfo{volume}{25}}, \bibinfo{pages}{49} (\bibinfo{year}{2005}).

\bibitem[{\citenamefont{Hofmann and M\"unzenberg}(2000)}]{SHIPReview}
\bibinfo{author}{\bibfnamefont{S.}~\bibnamefont{Hofmann}} \bibnamefont{and}
  \bibinfo{author}{\bibfnamefont{G.}~\bibnamefont{M\"unzenberg}},
  \bibinfo{journal}{Rev. Mod. Phys.} \textbf{\bibinfo{volume}{72}},
  \bibinfo{pages}{733} (\bibinfo{year}{2000}).

\bibitem[{\citenamefont{Neumayr et~al.}(2006)\citenamefont{Neumayr 
  et~al.}}]{Juergen_GasCell}
\bibinfo{author}{\bibfnamefont{J.~B.}~\bibnamefont{Neumayr}}
  \bibnamefont{et~al.},
  \bibinfo{journal}{Nucl. Instrum. and Methods B} \textbf{\bibinfo{volume}{244}},
  \bibinfo{pages}{489} (\bibinfo{year}{2006}).

\bibitem[{\citenamefont{Herfurth}(2003)}]{Frank_Buncher}
\bibinfo{author}{\bibfnamefont{F.}~\bibnamefont{Herfurth}},
  \bibinfo{journal}{Nucl. Instr. and Meth. B} \textbf{\bibinfo{volume}{204}},
  \bibinfo{pages}{587} (\bibinfo{year}{2003}).

\bibitem[{\citenamefont{Chaudhuri}(2006)}]{Ankur_Paper}
\bibinfo{author}{\bibfnamefont{A.}~\bibnamefont{Chaudhuri}},
  \bibinfo{journal}{Eur. Phys. J. A}  (\bibinfo{note}{to
  be published}).

\bibitem[{\citenamefont{Savard et~al.}(1991)\citenamefont{Savard et~al.}}]{Savard_BufferGasCooling}
\bibinfo{author}{\bibfnamefont{G.}~\bibnamefont{Savard}}
  \bibnamefont{et~al.},
  \bibinfo{journal}{Phys. Lett. A} \textbf{\bibinfo{volume}{158}},
  \bibinfo{pages}{247} (\bibinfo{year}{1991}).

\bibitem[{\citenamefont{Gr\"aff et~al.}(1980)\citenamefont{Gr\"aff, Kalinowsky,
  and Traut}}]{Graeff_TOFICR}
\bibinfo{author}{\bibfnamefont{G.}~\bibnamefont{Gr\"aff}},
  \bibinfo{author}{\bibfnamefont{H.}~\bibnamefont{Kalinowsky}},
  \bibnamefont{and} \bibinfo{author}{\bibfnamefont{J.}~\bibnamefont{Traut}},
  \bibinfo{journal}{Z. Phys. A} \textbf{\bibinfo{volume}{297}},
  \bibinfo{pages}{35} (\bibinfo{year}{1980}).

\bibitem[{\citenamefont{Block}(2006)}]{Michael_NIMPaper}
\bibinfo{author}{\bibfnamefont{M.}~\bibnamefont{Block}},
  \bibinfo{journal}{Nucl. Instrum. Methods B}  (\bibinfo{note}{to be published}).

\bibitem[{\citenamefont{Audi et~al.}(2003)\citenamefont{Audi, Bersillon,
  Blachot, Wapstra, and Thibault}}]{AME2003}
\bibinfo{author}{\bibfnamefont{G.}~\bibnamefont{Audi}},
  \bibinfo{author}{\bibfnamefont{O.}~\bibnamefont{Bersillon}},
  \bibinfo{author}{\bibfnamefont{J.}~\bibnamefont{Blachot}},
  \bibinfo{author}{\bibfnamefont{A.~H.}~\bibnamefont{Wapstra}},
  \bibnamefont{and} \bibinfo{author}{\bibfnamefont{C.}~\bibnamefont{Thibault}},
  \bibinfo{journal}{Nucl. Phys. A} \textbf{\bibinfo{volume}{729}},
  \bibinfo{pages}{3} (\bibinfo{year}{2003}).

\bibitem[{\citenamefont{Litvinov et~al.}(2005)\citenamefont{Litvinov
  et~al.}}]{Litvinov_ESRMeas}
\bibinfo{author}{\bibfnamefont{Y.~A.}~\bibnamefont{Litvinov}}
  \bibnamefont{et~al.}, \bibinfo{journal}{Nucl. Phys. A}
  \textbf{\bibinfo{volume}{756}}, \bibinfo{pages}{3} (\bibinfo{year}{2005}).

\bibitem[{\citenamefont{Beck et~al.}(2000)\citenamefont{Beck
  et~al.}}]{Didi_ISOLTRAP}
\bibinfo{author}{\bibfnamefont{D.}~\bibnamefont{Beck}}
  \bibnamefont{et~al.}, \bibinfo{journal}{Eur. Phys. J. A}
  \textbf{\bibinfo{volume}{8}}, \bibinfo{pages}{307} (\bibinfo{year}{2000}).

\bibitem[{\citenamefont{Rauth}(2006)}]{MyEPJAPaper}
\bibinfo{author}{\bibfnamefont{C.}~\bibnamefont{Rauth}}, \bibinfo{journal}{Eur.
  Phys. J. A}  (\bibinfo{note}{to be published}).

\bibitem[{\citenamefont{Seweryniak et~al.}(1997)\citenamefont{Seweryniak
  et~al.}}]{Seweryniak_Tm147}
\bibinfo{author}{\bibfnamefont{D.}~\bibnamefont{Seweryniak}}
  \bibnamefont{et~al.}, \bibinfo{journal}{Phys. Rev. C}
  \textbf{\bibinfo{volume}{55}}, \bibinfo{pages}{R2137} (\bibinfo{year}{1997}).

\bibitem[{\citenamefont{Sellin et~al.}(1993)\citenamefont{Sellin et~al.}}]{Sellin_Tm147}
\bibinfo{author}{\bibfnamefont{P.~J.}~\bibnamefont{Sellin}}
  \bibnamefont{et~al.},
  \bibinfo{journal}{Phys. Rev. C} \textbf{\bibinfo{volume}{47}},
  \bibinfo{pages}{1933} (\bibinfo{year}{1993}).

\bibitem[{\citenamefont{Hofmann}(1995)}]{Sigurd_ProtonRadioactivity}
\bibinfo{author}{\bibfnamefont{S.}~\bibnamefont{Hofmann}},
  \bibinfo{journal}{Radiochim. Acta} \textbf{\bibinfo{volume}{70/71}},
  \bibinfo{pages}{93} (\bibinfo{year}{1995}).

\end{thebibliography}
%
%

%
\end{document}